\documentclass[prb,aps,showpacs,superscriptaddress,jpgfig,twocolumn,floatfix]{revtex4-1}
\usepackage{graphicx}
\usepackage{dcolumn}

\begin{document}
\title {Conductivity noise study of the insulator-metal transition and phase co-existence in epitaxial samarium nickelate thin films}
\author{Anindita Sahoo}
\address{Department of Physics, Indian Institute of Science, Bangalore 560 012, India}
\author{Sieu D. Ha}
\author{Shriram Ramanathan}
\address{School of Engineering and Applied Sciences, Harvard University, Cambridge, MA 02138, USA}
\author{Arindam Ghosh}
\address{Department of Physics, Indian Institute of Science, Bangalore 560 012, India}
\date{\today}

\begin{abstract}
Interaction between the lattice and the orbital degrees of freedom not only makes rare-earth nickelates unusually ``bad metal", but also introduces a temperature driven insulator-metal phase transition. Here we investigate this insulator-metal phase transition in thin films of $\mathrm{SmNiO_3}$ using the slow time dependent fluctuations (noise) in resistivity. The normalized magnitude of noise is found to be extremely large, being nearly eight orders of magnitude higher than thin films of common disordered metallic systems, and indicates electrical conduction via classical percolation in a spatially inhomogeneous medium. The higher order statistics of the fluctuations indicate a strong non-Gaussian component of noise close to the transition, attributing the inhomogeneity to co-existence of the metallic and insulating phases. Our experiment offers a new insight on the impact of lattice-orbital coupling on the microscopic mechanism of electron transport in the rare-earth nickelates.
\end{abstract}

\maketitle

The interest in Perovskite nickelates ($R\mathrm{NiO_3}$, $R =$ rare earth elements) originates from a rich phase space that exhibits many interesting phenomena including sharp insulator-metal transition, charge or orbital ordering, magnetic ordering etc.~\cite{perovskite_review,perovskite_paper18,perovskite_paper4,perovskite_paper8,perovskite_paper9}. The radius of the rare earth element controls the bond lengths as well as the tilt of the $(NiO_6)^{3-}$ octahedra, and hence the orbital overlap which allows a large variation in the electron interaction within the $R\mathrm{NiO_3}$ family~\cite{perovskite_paper11,perovskite_paper9}. The correlation effects are however complicated by the structural uniqueness of these compounds. For small radius, e.g. $R = $ Sm, Eu, Gd,..., Lu, an antiferromagentic insulator (AFI) ground state is stabilized along with a monoclinic distortion of the crystal structure at low temperatures ($T$). With increasing $T$, two successive phase transitions occur, (i) AFI to PI (paramagnetic insulator) magnetic transition at the Neel temperature $T_N$, and (ii) PI to PM (paramagnetic metal) transition at $T_{IM}$ associated with structural transition from the monoclinic ($P2_1/n$) to orthorhombic ($Pbnm$) symmetries~\cite{perovskite_paper11,perovskite_paper14,perovskite_paper15}. Due to the complex interplay between the structural distortions and Ni$-$O orbital interactions, the nature of the AFI-PI and PI-PM transitions remains poorly understood, in spite of many experimental investigations in the past~\cite{shriram_nat,raman_at_mit2,susceptibility,neel_temp_from_r,diffraction}.

$\mathrm{SmNiO_3}$ is unique in the sense that it exists in the intermediate regime between strong and weak inter-electron interaction regimes, and displays the highest $T_N$, in the rare earth nickelate family. For smaller ionic radii, {\it i.e.} for $R =$ Sm, Eu, Gd, Dy, etc, where $T_N < T_{IM}$, the PI-PM transition is suggested to be first order in nature~\cite{perovskite_review,so_transition1}, although the microscopic details of this first order phase transition remains obscured. For $\mathrm{SmNiO_3}$, the only evidence so far for a structural transition at $T_{IM}$ is an indirect Raman comparison with $\mathrm{NdNiO_3}$ (Ref~\cite{raman_at_mit2}). While the conventional time-averaged electrical resistivity measurements can indicate an overall change in the nature of electronic states, {\it i.e.} from an insulator to a metal, alternative experimental probes are required to explore the transition kinetics and other related issues, such as the microscopic details of electron transport close to the PI-PM transition, which can impact the application of these systems as switches~\cite{nickelate_transistor}.

The low-frequency fluctuations in electrical resistivity, often known as the $1/f$ or flicker noise, provides a powerful tool to probe the coupling of electronic and structural degrees of freedom in these materials. The magnitude of noise is the sum of cross-terms in current autocorrelation function, originating from different sub-volumes of the material. Each of these terms are proportional to the square of the local current density (Cohn's theorem~\cite{cohn}), making the noise power proportional to the fourth power of the local current density. Consequently, the noise spectrum is extremely sensitive not only to the extent of structural disorder, but also to the slow time-dependent kinetics and rearrangements of the disorder. This has been verified on numerous occasions with metal films~\cite{dutta_horn_noise,noise_metal,weissman_rmp}, doped semiconductors~\cite{swastik_prl,arindam_prl}, shape memory alloys~\cite{Chandni_apl,Chandni_actamat} and so on, over the past several decades. In this work, we have probed for the first time the insulator to metal phase transitions in thin epitaxial films of $\mathrm{SmNiO_3}$ on crystalline LaAlO$_3$ substrates, with the slow time-dependent fluctuations, or noise, in the electrical resistivity to capture the stochasticity of the transformation kinetics during the phase transitions. We find, (a) extremely large noise magnitude in these systems that follow the kinetics of classical percolation in an inhomogeneous medium, and (b) a strong non-Gaussian component in the resistivity fluctuations close to the PI-PM transition. Both results suggest an intriguing co-existence of the parent and product phases during the insulator-metal electronic transition, which we attribute to the simultaneous structural change.

\begin{figure}
\centering
\includegraphics[width=1\linewidth]{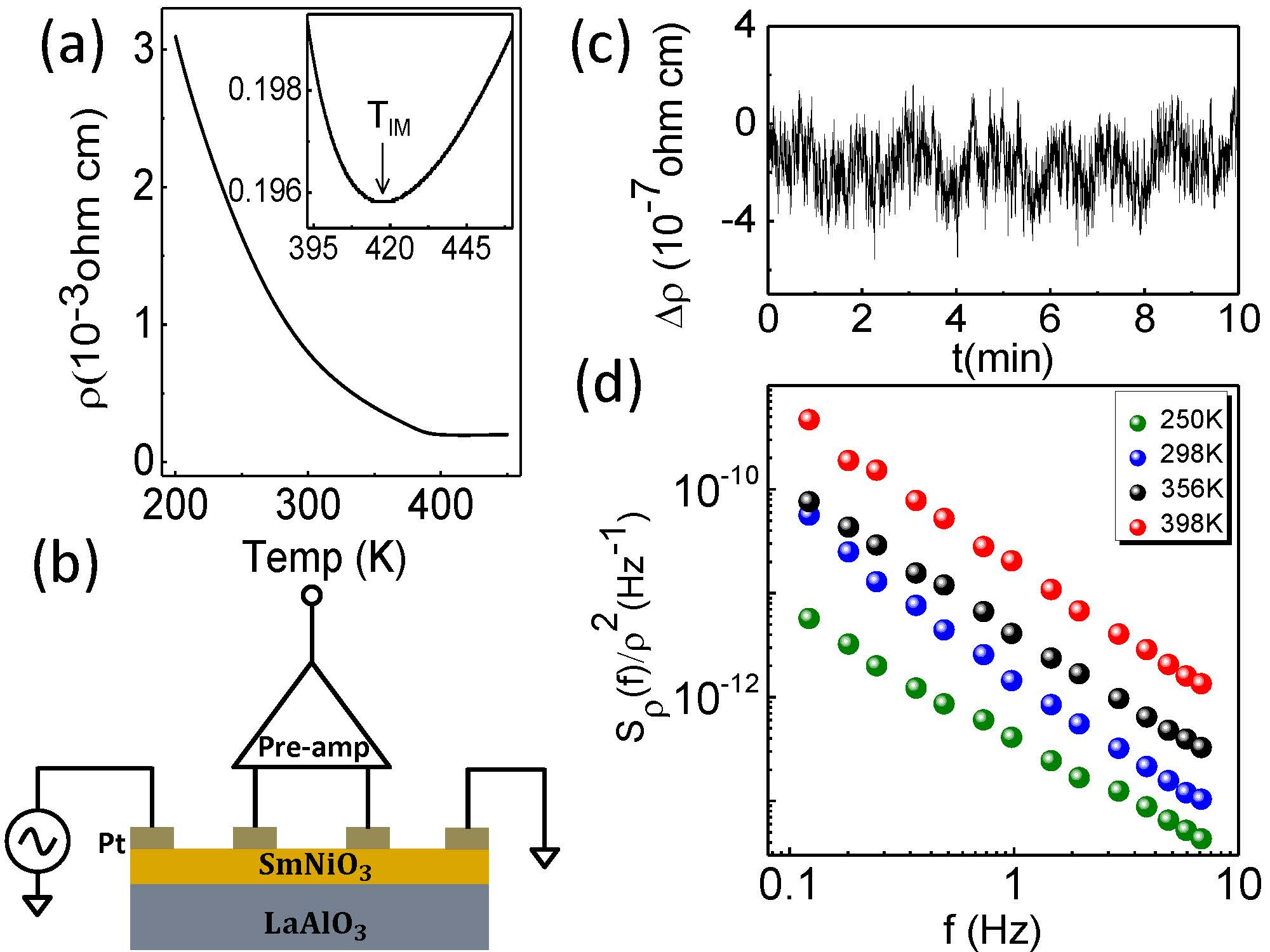}
\caption{(a) Resistivity ($\rho$) vs temperature plot of $\mathrm{SmNiO_3}$ thin film which displays insulator to metal transition around 418K (magnified in the inset). (b) Schematic of the device for resistivity and noise measurement. (c) Typical resistivity fluctuation with time at a particular temperature.(d) PSD at different temperatures showing $1/f$ noise.}
\label{Fig1}
\end{figure}

In our experiments here we used thin ($\approx 80$~nm) films of epitaxial  deposited on crystalline $\mathrm{LaAlO_3}$ substrate by rf magnetron sputtering in high background pressure~\cite{sample_preparation}.  For electrical measurements, 100~nm platinum was deposited on $\mathrm{SmNiO_3}$ thin film by dc sputtering for low resistance contact pads. The temperature dependence of resistivity ($\rho$) of a typical $\mathrm{SmNiO_3}$ film is shown in Fig.~1a which clearly identifies the metal-insulator transition at $T_{IM} \approx 418$~K (inset in FIG.~1a). No hysteresis in $\rho$ could be detected in repeated temperature sweeps between 200K and 450K, at a rate as high as 1K/min~\cite{so_transition3,susceptibility}. The schematic of the device as well as the electrical circuit to measure resistivity and noise are displayed in Fig.~1b. Fig.~1c,d outlines the noise characteristics and the power spectral density (PSD) of noise (see supplementary online material for detailed experimental procedure). A typical time series of resistivity fluctuations $\Delta\rho(t)$ at the fixed temperature $T$ ($\approx 387$~K) is shown in Fig.~1c. The PSD $S_\rho(f) \propto 1/f^\alpha$, with $\alpha \sim 1$ represents the $1/f$ noise in resistivity. Fig.~1d illustrates the experimental PSD, normalized by the square of the corresponding resistivity at different $T$, which also shows that the magnitude of noise increases with increasing $T$, but its spectral content {\it i.e.} the $1/f$ nature with $\alpha \sim 1 - 1.3$, remains intact (Fig~2b).

\begin{figure}
\centering
\includegraphics[width=1\linewidth]{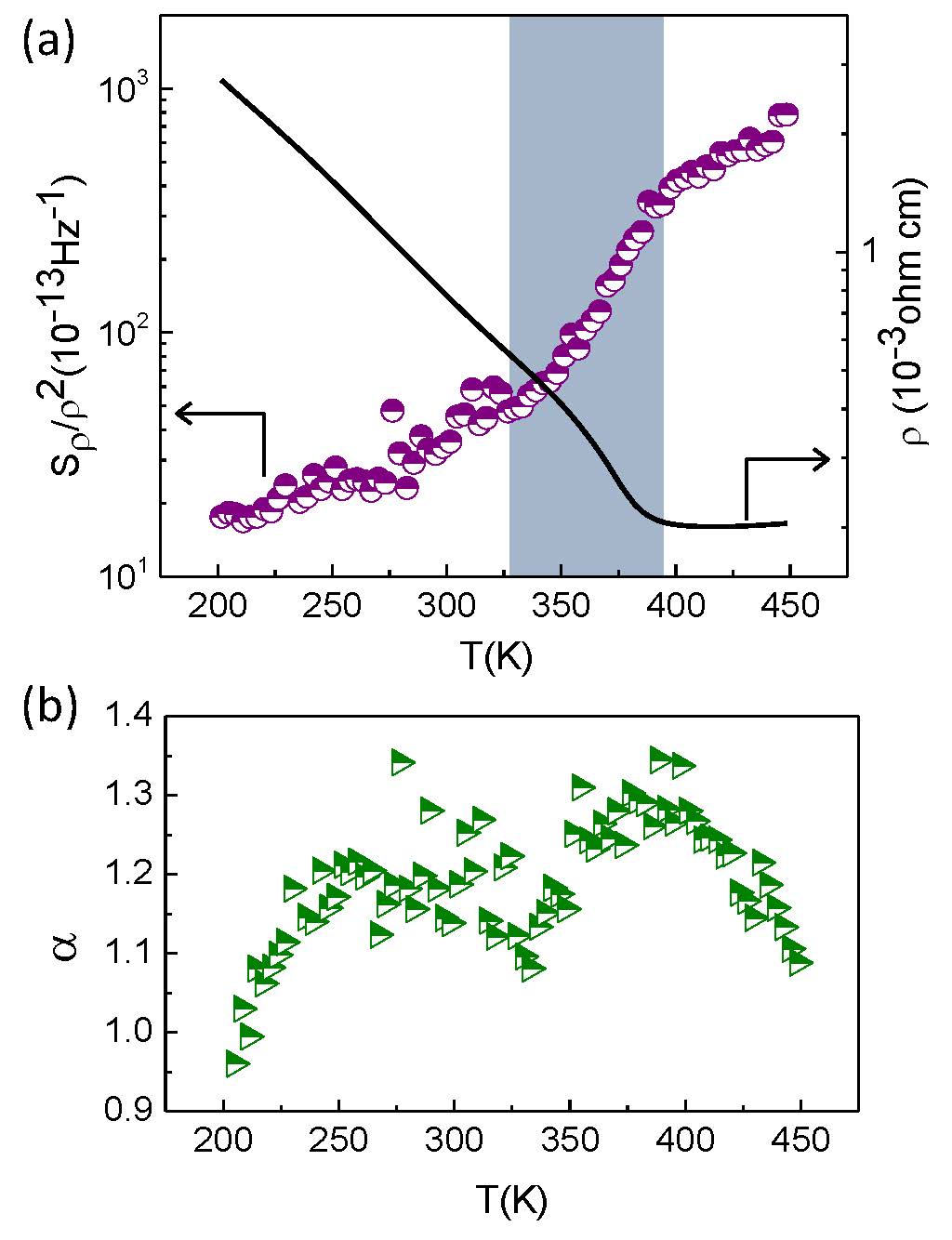}
\caption{ (a) Sample resistivity ($\rho$) together with the corresponding PSD at 1Hz ($S_\rho/\rho^2$) which increases monotonically with increasing temperature. The blue background represents percolative transport regime ($\approx 330-390$~K). (b) The exponent $\alpha$ with increasing temperature where $\alpha$ ranges from 1.1 to 1.3. }
\label{Fig2}
\end{figure}

To estimate the noise magnitude across the PI-PM phase transitions in $\mathrm{SmNiO_3}$, we measured the PSD of the fluctuations in $\rho$ while increasing $T$ from $200$~K to $450$~K. In Fig.~2a, we find that the normalized $S_\rho/\rho^2$ increases monotonically with temperature over the entire temperature range, but the $T$-dependence weakens considerably at the onset of the metallic behaviour. In disordered metals, thermally activated relaxation of disorder such as defect cluster, leads to a peak in noise at temperature scale characteristic to defect activation energy~\cite{dutta_horn_noise,mantese_noise}, although we did not observe any peak even at the highest experimental $T$. Alternatively, the weakening of $T$-dependence noise may also occur due to reduced sensitivity of $\rho$ to disorder relaxation in the metallic regime, compared to the insulating phase~\cite{Chandni_apl}. Nevertheless, the change in the temperature dependence in noise magnitude can be a qualitative marker to the onset of the metallic regime.

To obtain a deeper understanding on the origin of noise close to the transition, and to compare its magnitude to other conducting systems, we have normalized the noise magnitude to the so-called Hooge parameter $\gamma_H$ as below:

\begin{equation}
\label{hooge}
\gamma_H = n\Omega\frac{fS_\rho}{\rho^2}
\end{equation}

\noindent where $n$, and $\Omega$ are the density of charge carriers and the volume of the film between the voltage contacts, respectively (details of the estimation of carrier density is given in Ref~\cite{perovskite_paper8}). In Fig.~3, we have plotted the estimated $\gamma_H$ at $f = 1$~Hz as a function of $\rho$ of the film for pristine as well as aged device which had undergone a large number ($\sim 50$) of thermal cycles. The resistivity at the PI-PM transition corresponds to the left of the panel (dashed vertical lines), where $\gamma_H$ is large. $\gamma_H$ decreases rapidly as the system is driven into the insulating phase (high-$\rho$ regime). We note two striking features in Fig.~3: First, the experimental magnitude of $\gamma_H \sim 10^3 - 10^5$ exceeds the conventional metallic films by $6 - 8$ orders of magnitude, when the film approaches the PI-PM transition. This immediately suggests that the microstructural details of these films are very different from the simple disordered metallic or quasi-metallic systems in spite of comparable magnitude of $\rho$ . Early noise experiments in metallic $\mathrm{LaNiO_3}$ films also indicated a large $\gamma_H$ ($\sim 25$), arising from long-range diffusion of oxygen ions that resulted in $\alpha \rightarrow 1.5$ at higher $T$ (Ref~\cite{arindam_lanio3}). No such trend in $\alpha ~(\sim 1 - 1.3)$ was observed for the $\mathrm{SmNiO_3}$ films as shown in Fig.~2b, indicating noise to arise from a different physical mechanism. Second, we find $\gamma_H \propto \rho^{-w}$, with $w = 2.0 \pm 0.1$ for the pristine device, as the PI-PM transition is approached ($\rho \lesssim 0.5$~mohm-cm and $T \gtrsim 330$~K), which is strongly indicative of `random void model' associated with the classical percolation of electrons through a spatially inhomogeneous medium~\cite{percolation}. Deep inside the insulating regime ($\rho > 0.5$~mohm-cm), the dependence of $\gamma_H$ on $\rho$ changes, and $w$ reduces to $\approx 0.8 \pm 0.1$. We note here that, (1) disordered thin films of high-$T_c$ oxide also show very similar behaviour which was attributed to classical percolation~\cite{kiss2}, and (2) such behaviour extends to the metal-insulator transition point when the film is aged, usually indicative of strong morphological or stoichiometric inhomogeneity probably via loss of oxygen.

\begin{figure}
\centering
\includegraphics[width=1\linewidth]{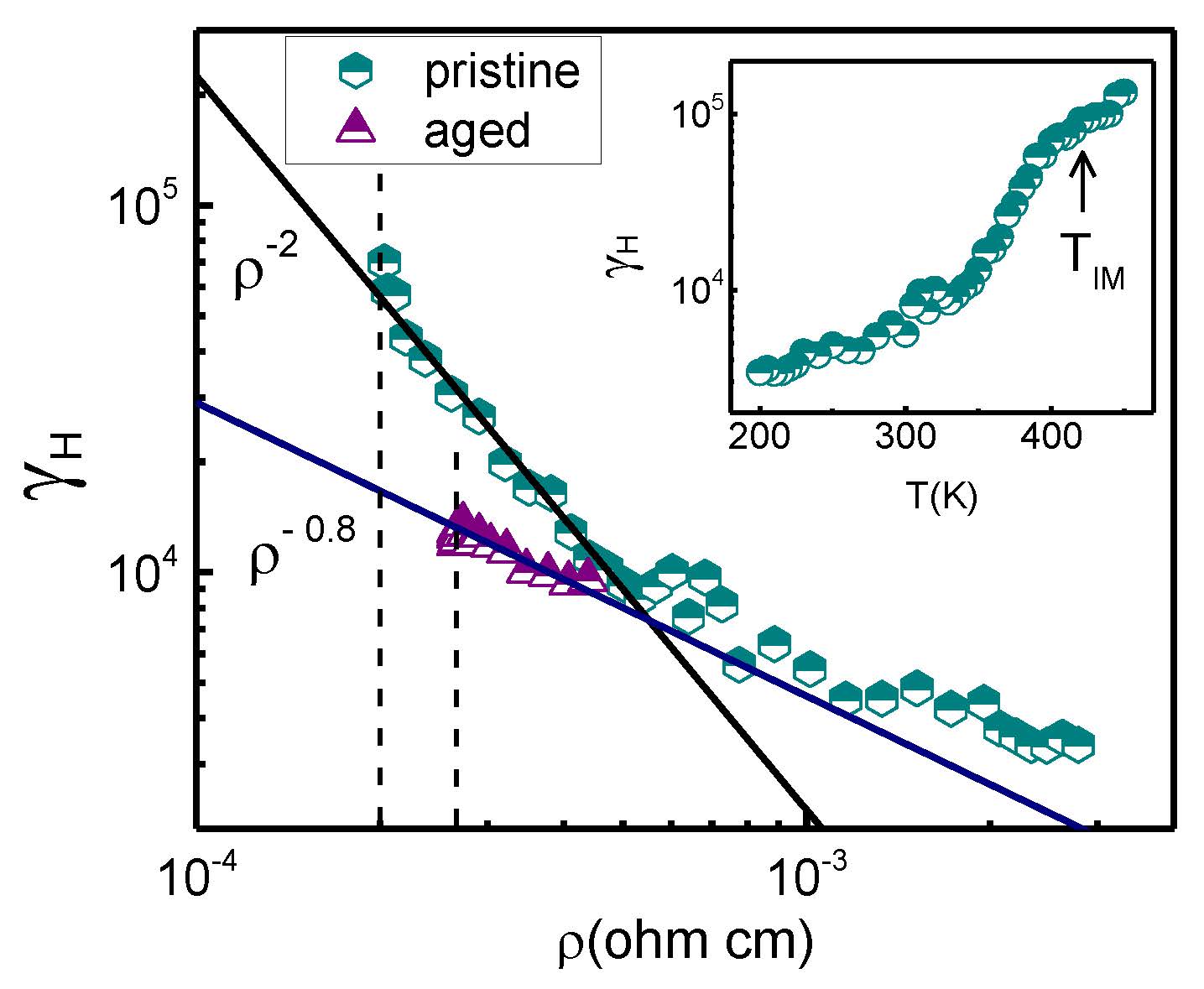}
\caption{ Hooge parameter $\gamma_H$ as a function of resistivity $\rho$ which increases rapidly as the system is driven to metallic phase of lower resistivity with increasing temperature for both pristine \& aged devices. The black and blue lines represent the linear fits which displays the $\rho^{-2}$ and $\rho^{-0.8}$ dependence of $\gamma_H$ respectively. The vertical dashed lines indicate the resistivities at $T_{IM}$ for both the devices. The inset shows the temperature dependence of $\gamma_H$ of the pristine device.}
\label{Fig3}
\end{figure}

A percolative transport also explains occurrence of the large $\gamma_H$ in our films. Noise studies at the electronic, magnetic, or structural phase transitions in numerous systems ranging from perovskite manganites~\cite{manganite_podzorov2}, high$-T_c$ superconducting films~\cite{kiss2}, organic semiconductors~\cite{organicsemi1,organicsemi2}, to shape memory alloys~\cite{Chandni_prb,Chandni_apl,Chandni_prl,Chandni_actamat}, have indicated very large $\gamma_H$ compared to common disordered films, arising from percolative kinetics. Close to the transition these systems contain two competing phases with unequal electrical resistivities that form a percolative network, which in our case can be readily identified as the metallic (high $T$) and the insulating (low $T$) phases corresponding to the orthorhombic and monoclinic structural symmetries, respectively. In Fig~2a we have shown the percolative transport regime (blue shaded background) where the temperature dependences of noise ($S_\rho/\rho^2$) as well as resistivity ($\rho$) differ from that in both insulating and metallic regimes. While the possibility of phase co-existence close to the PI-PM transition in the $\mathrm{SmNiO_3}$ films has been suggested before~\cite{raman_at_mit2}, the noise experiments here reveal the nature of electrical transport in this regime for the first time.

We now focus on the higher order statistics of resistivity fluctuations, and extract the non-Gaussian component from the measured fluctuations, which is known to be extremely sensitive to the thermodynamic details of the transition process~\cite{weissman_spinglass,swastik_prl,Chandni_prb}. Our objective here is to evaluate the `second spectrum' $S^{(2)}_{f_1}(f_2)$, defined as

\begin{equation}
S^{(2)}_{f_1}(f_2)=\frac{\lbrack \int_0^\infty \langle\Delta\rho^2(t)\Delta\rho^2(t+\tau)\rangle_t \cos(2\pi f_2\tau)\,d\tau \rbrack}{\lbrack \int_{f_L}^{f_H} S_\rho (f_1)\,df_1 \rbrack ^2}
\end{equation}

\noindent where $f_1$ and $f_2$ refer to the spectral frequencies of first spectrum and second spectrum respectively. Effectively, $S^{(2)}_{f_1}(f_2)$ is the power spectrum obtained from a time series of $\Delta\rho^2 (t)$, where $\Delta\rho^2(t)$ is evaluated from the PSD integrated over an octave around $f_1$. The details of the method is described in the supplementary online material and also published elsewhere~\cite{Chandni_prb,restle,second_spectrum}. In our experiments, each second spectrum was computed from a time series of $\Delta\rho^2(t)$, with each element in the time series evaluated from a frequency octave of $0.2$~Hz--$0.4$~Hz centred around $f_1=0.3$~Hz where the background noise floor is much lower than the sample noise. Fig.~4a shows the normalized $S^{(2)}_{f_1}(f_2)$ of noise of the sample at different temperatures and the second spectrum was generally weakly frequency-dependent irrespective of temperature. Close to $T_{IM}$ it exceeds the expected Gaussian background by several orders and this indicates emergence of strong non-Gaussian component of resistivity fluctuation at the PI-PM transition.

\begin{figure}
\centering
\includegraphics[width=1\linewidth]{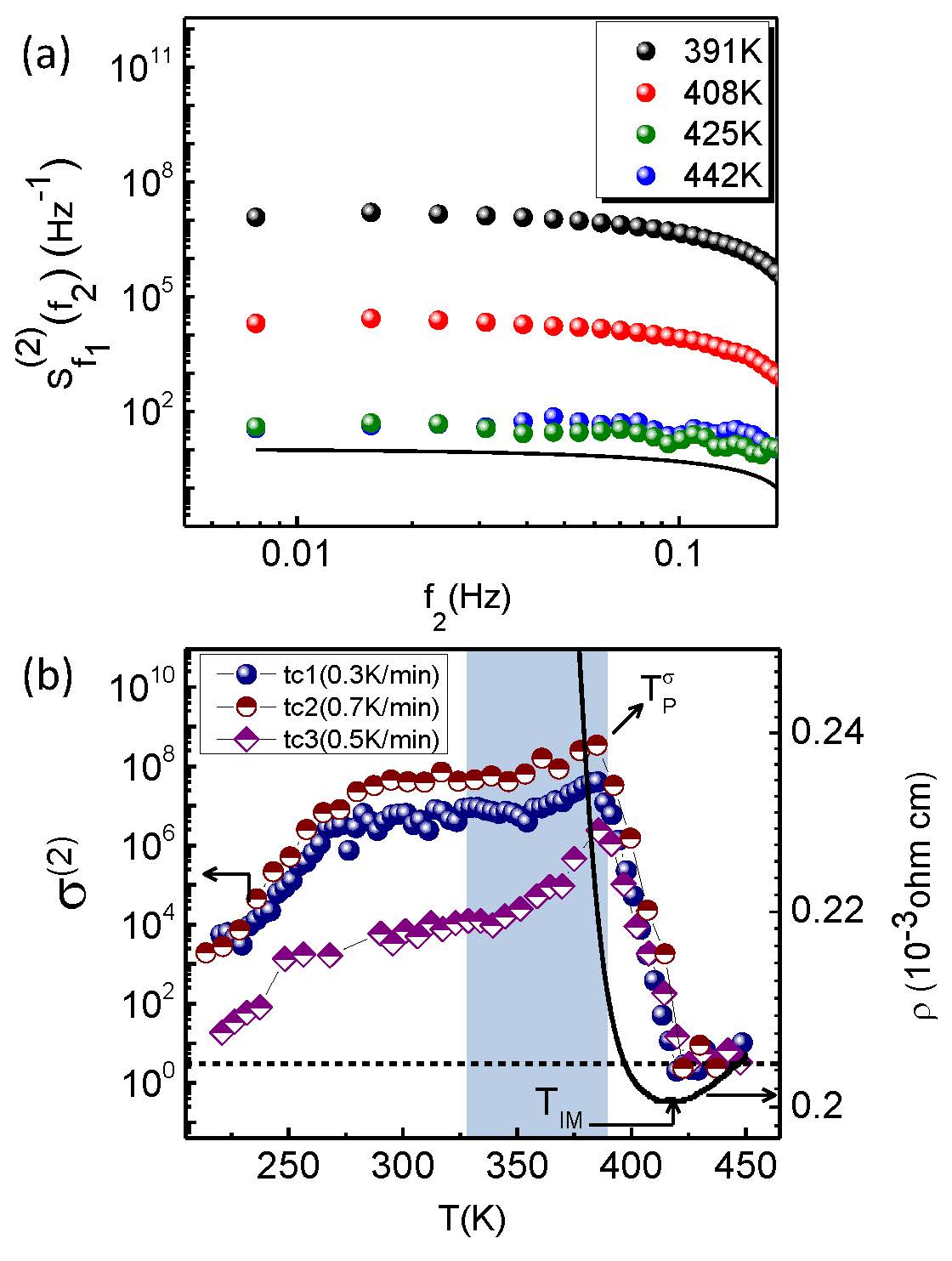}
\caption{ (a) Normalized second spectrum $S^{(2)}_{f_1}(f_2)$ at different temperatures which shows frequency independent white-noise characteristics. The solid line indicates the expected normalized Gaussian background~\cite{second_spectrum}. (b) Normalized variance of second spectrum, $\sigma^{(2)}$ with increasing temperature for different thermal cycles (tc) driven at different ramping rates 0.3K/min, 0.5K/min and 0.7K/min. The dashed line indicates the Gaussian background. The blue shaded background corresponds to the percolative regime identified from the noise data in Fig~3. The solid line represents the $\rho$ vs $T$ data which indicates insulator-metal transition at $T_{IM} \approx 418~K$.}
\label{Fig4}
\end{figure}

Both frequency and temperature dependence of the second spectrum provide crucial insight into the electronic transport during micro-structural evolution of the $\mathrm{SmNiO_3}$ films at the PI-PM transition. The second spectrum serves as an estimate of the non-Gaussian component of the resistivity fluctuation and strong non-Gaussian component indicates two possibilities: (i) long range correlation between fluctuators associated with the atomic avalanches which also results in frequency-dependent second spectrum ($S^{(2)}_{f_1}(f_2)$) (Ref~\cite{weissman_spinglass,Chandni_prl}) and (ii) co-existance of phases with unequal resistivities which gives rise to a percolative network. In the latter case, metallic element of the network having lower resistivity carries larger current whereas the insulating element in parallel having larger resistivity carries lower current which is known as dynamical current redistribution (DCR)~\cite{dynamical_current_redistribution}. Although the fluctuators are statistically independent resulting in near-white second spectrum, large local resistivity fluctuations in the system due to DCR leads to a strong non-Gaussian component. Based on these, we highlight two implications of the results shown in Fig~4: (1) The nearly frequency independent $S^{(2)}_{f_1}(f_2)$ (Fig~4a) provides evidence against long range interaction, and makes DCR a more likely mechanism for non-Gaussianity. This is also supported by $\gamma_H$ vs $\rho$ data in Fig~3 which gives a direct evidence of classical percolation. (2) For a physically intuitive comparison, we integrated $S^{(2)}_{f_1}(f_2)$ within the experimental bandwidth, so that $\sigma^{(2)}=\int_{0}^{f_H-f_L} S^{(2)}_{f_1}(f_2)\,df_2$ is proportional to the kurtosis of the fluctuations in $\rho$. For Gaussian resistivity fluctuations, one expects $\sigma^{(2)}\approx 3$ (Ref~\cite{koushik_prl}) indicated by the dashed line in Fig~4b. We have plotted $\sigma^{(2)}$ as a function of $T$ in Fig~4b for three heating cycles driven at different temperature sweep rates. Deep inside the insulating regime, $\sigma^{(2)}$ approaches Gaussianity for $T\lesssim 200$~K, although its magnitude depends on the thermal history and varies from one thermal cycle to the other without any specific trend. With increasing $T$, $\sigma^{(2)}$ increases nearly exponentially, and reaches a maximum value of $\sim 10^6-10^8$ at $T\sim 385$~K, irrespective of temperature ramping rates. Above this temperature, $\sigma^{(2)}$ collapses dramatically and acquires its Gaussian magnitude (dashed line) at $T\simeq 420$~K, which is essentially the temperature scale $T_{IM}$ for the onset of metallic transport in the film (see $\rho-T$ data as the solid line in Fig~4b, which is replotted from Fig~1a).

The unique $T$-variation of $\sigma^{(2)}$ across the PI-PM transition in the $\mathrm{SmNiO_3}$ films, reveals several intriguing features. For $T\lesssim 385$~K the non-Gaussianity decreases with decreasing $T$, albeit rather slowly with $T$. This is readily attributed to the increasing fraction of the insulating phase, nucleating within the metallic matrix. At high $T$ ($>385$~K), the metallic domains in the percolative network starts expanding with increasing $T$ resulting in reduction of the effect of DCR. The magnitude of $\sigma^{(2)}$ decreases abruptly by nearly eight orders of magnitude, suggestive of a first order phase transition, and the sample becomes mostly metallic within a temperature range of $30-35$~K. For $T>T_{IM}$ ($\sim 418$~K) the absence of DCR leads to Gaussian resistivity fluctuations. Earlier studies of calorimetry and linear thermal expansion~\cite{so_transition1} and variable temperature neutron diffraction~\cite{struc_change_below_im_trans_temp} on $\mathrm{SmNiO_3}$ also confirmed the co-existence of parent and product phases during the transition period and revealed that structural changes during insulator-metal transition is a smooth process, although these technique could detect structural changes only down to $\sim 20-30$~K below $T_{IM}$. In our case, the maximum in $\sigma^{(2)}$ at $T^\sigma_P$ ($\sim 385$~K) represents the percolation threshold, separating the metal-dominated ($T>T^\sigma_P$) and insulator-dominated ($T<T^\sigma_P$) regions. Finally, the observation that $T^\sigma_P$ remains essentially unchanged irrespective of temperature driving rate indicates that contribution of athermal atomic avalanches during the PI-PM transition to be undetectably small, as supported by the white second spectra of Fig~4a~\cite{Chandni_prb,Chandni_prl}.

In conclusion, the extremely large magnitude of resistivity noise indicates the presence of percolative electron transport in $\mathrm{SmNiO_3}$ thin film at the insulator-metal phase transition. On the insulating side of the transition, the noise develops a strong non-Gaussian component that is shown to arise from dynamical current redistribution in a percolative network, thereby providing additional confirmation of phase co-existence. The phase co-existence disappears abruptly at the onset of metallic transport, indicating its origin is linked to the lattice-orbital coupling in epitaxial $\mathrm{SmNiO_3}$ thin films. The approach presented here could also be extended to studies of phase transition in other correlated oxide system.

AS and AG acknowledge the Department of Science and Technology (DST) for a funded project. SDH and SR acknowledge ARO MURI for financial support.

\end{document}